\pdfoutput=1
\documentclass{article}

\usepackage{arxiv}

\usepackage[utf8]{inputenc} 
\usepackage[T1]{fontenc}    
\usepackage{hyperref}       
\usepackage{url}            
\usepackage{booktabs}       
\usepackage{amsfonts}       
\usepackage{nicefrac}       
\usepackage{microtype}      
\usepackage{lipsum}
\usepackage{graphicx}
\graphicspath{ {./images/} }

\usepackage[version=3]{mhchem} 

\usepackage{natbib}
\setcitestyle{numbers}
\setcitestyle{square}
\usepackage{xcolor}

\title{Accreting Primordial Black Holes as Dark Matter Constituents}

\author{
T. Kenneth Fowler \\
  Department of Nuclear Engineering\\
 University of California at Berkeley\\
4153 Etcheverry Hall\\
  Berkeley, CA, 94720 USA\\
  \texttt{TKFowler5@aol.com} \\
   \And
Richard Anantua \\
  Department of Physics and Astronomy\\
  The University of Texas at San Antonio \\
  One UTSA Circle
  \\San Antonio, TX 78249, USA\ \\
  \texttt{richard.anantua@utsa.edu} \\
}

\begin{document}
\maketitle
$\qquad$
\newline
$\qquad$
\newline
\begin{abstract}
We show how magnetic accretion of 
positronium (
electron-positron) 
 plasma by primordial 
black holes
might 
significantly contribute to the mass of dark matter in the present Universe. 
Assuming that background gamma radiation is primordial black hole Hawking radiation rules out Bondi accretion, while magnetic accretion known from studies of active galactic nuclei could explain the abundance of dark matter. Various accretion scenarios are discussed. 
\newline
$\qquad$
\end{abstract}

$\qquad$
\newline
$\qquad$
\newline
$\qquad$
\newline
\textbf{{1. Introduction}} \\
\indent As reviewed in \cite{carr2021constraints}, many authors have pursued Hawking’s early suggestion
that primordial black holes might have contributed significantly to dark matter.
Here we focus on the period 0.01 s to 14 s into the Big Bang, when relativistic
positronium plasma was a dominant constituent of the mass, with $10^{9}$ electrons and
positrons per proton and neutron \cite{weinberg1979first}. Then accreting only a tiny fraction of
primordial positronium preserved as black holes could account for dark matter
mass $M_{DM}$.
This paper makes three main points: (a) assuming background gamma radiation is
Hawking radiation from dark matter fixes $M^{15-18}$
g  
as the range of black hole
masses; (b) this mass range is accessible by accretion by magnetic
turbulence, similar to that identified in general relativistic MHD simulations of
active galactic nuclei (AGNs) \cite{narayan2012grmhd, mckinney2012general, davis2020magnetohydrodynamics}; and (c) transport by magnetic turbulence can explain dark matter abundance. \\


\vspace{-3mm}
\textbf{2. Hawking Radiation and Accretion Scenarios}\\
\indent Primordial black holes are theorized to be surrounded by a thermal bath of radiated particles whose temperature is inversely proportional to black hole mass \cite{hawking1974hawkingradiation}.  Such  particles range from electromagnetic rays across the spectrum to gravitons \cite{Anantua2009PBHGraviton}. Among radiations not fully understood is the gamma ray background,
extending from keV's to 10 MeV and more, cited as possibly being of
primordial black hole origin \cite{carr2021constraints}. In notes below, we obtain a mass distribution f(M) fitting Hawking radiation to background gammas. We set $\int_{M1}^{M2}dM Mf = M_{DM} = 6M_{0}$ (ordinary matter, current best estimate). We obtain $M_{2} \approx 10^{18}$ g while $M_{1} \approx
10^{15}$ g is the smallest not yet evaporated \cite{carr2021constraints}.

This mass range $10^{15-18}$ g places strong constraints on accretion scenarios. Standard Bondi accretion \cite{frank2002accretion}[Sect. 2.5] yields masses greater than the Sun \cite{carr2021constraints}. 
Only accretion sweeping up mass over a large radius can give smaller masses $M \approx M_{BONDI}$ ($\varrho_{amb}/\varrho$) for ambient density $\varrho_{amb}$ and density $\varrho$ piling up at the black hole. Examples are Shakura-Sunyaev viscous flow [8; Chap. 5] and, more importantly, magneto-rotational-instability (MRI) \cite{balbus1998instability} that turns out may 
explain dark matter abundance. \\

\textbf{{3. Dark Matter Abundance}}

Consider a poloidal magnetic seed field $(B_{\phi} = 0)$. MRI accretion geometry is determined by the evolving magnetic field, giving $V_{z}/V_{r}$ = $B_{z}/B_{\phi}$ \cite{colgate2014quasi}[Eqs. 12-14], giving in turn spherical flow $V_{z}/V_{r} \ge 1$ while $B_{\phi}$ is growing, true in the time available
in the expanding primordial plasma. A toroidal seed field behaves similarly \cite{mckinney2012general, balbus1998instability}. Using this argument, we consider accretion by an isolated black hole mass M
with accretion velocity $V_{r} = V_{MRI}$ everywhere on a sphere of expanding radius $r = R_{0}(t)$. Let $f^{*}$ be the fraction of mass accreted inside this sphere. We obtain:

\begin{subequations}
\begin{align}
    dM/dt \approx M/t \approx 4\pi\rho_{amb}R_{0}^{2}|V_{r}|\label{eq:1a}\\
    f^{*} \approx [4\pi\rho_{amb}R_{0}^{2}|V_{r}|t/((4\pi/3)\rho_{amb}R_{0}^{3}] \approx (3|V_{R}|t/R_{0})\label{eq:1b}\\
    |V_{r}| \approx (f^{*}/3)(R_{0}/t) \approx (\xi/r)^{2}(MG/R_{0})^{1/2}\label{eq:1c}\\
    R_{0}(t) \approx [0.021/(\xi/r)^{4/3}]M(t)^{1/3}t^{2/3}\label{eq:1d}\\   
    M_{DM}/M_{0} \approx [10^{9}/((m_{p} + m_{n})/m_{e})](3|V_{r}|t/R_{0})\label{eq:1e}
\end{align}
\end{subequations}

Equation (\ref{eq:1b}) uses Equation (\ref{eq:1a}) where here and hereafter t serves both as
dynamical elapsed time and as primordial time in ambient density $\rho_{amb} \approx 4 \times 10^{5}/t^{2}$ and temperature $T = T_{keV} = 1000/\sqrt{t}$ \cite{weinberg1979first}[pp. 4,5]. Equation (\ref{eq:1c}) equates $V_{r}$ from Equation (\ref{eq:1b}) to MRI on the far right hand side (in terms of magnetic field line
fluctuation ($\xi/r$)). Combining Equations (\ref{eq:1a}, \ref{eq:1c}) gives $R_{0}$ in Equation (\ref{eq:1d}). Combining
Equations (\ref{eq:1c}, \ref{eq:1d}, \ref{eq:1e}) gives the ratio of dark matter $M_{DM}$ to ordinary matter $M_{0}$, using $10^{9}$ electrons per proton and neutron cited above. Typically $(\xi/r) \approx 0.1$ \cite{colgate2015quasi}[Sect.
6.1], yielding the current best estimate $M_{DM} /M_{0} \approx 6$.

The assumption that M is isolated is marginally satisfied. The average
spacing between black holes $\Delta_{BH} \approx (1/n_{BH})^{1/3} = (<M>/f^{*}\rho_{amb})^{1/3} \approx R_{0}$ for $<M> \approx 10^{16}$ gm/cc at t=1 s. Acceleration between black holes with this spacing is small $(<M>G/\Delta_{BH}^{2})t \approx 10^{-7} (\Delta_{BH}/t))$.

Primordial conditions were right for MRI. The primordial magnetic field $\approx$ 1 gauss is sufficient to initiate MRI \cite{enqvist1998primordial}; and the Peebles mechanism could create the required rotation \cite{peebles1969origin}. Differences in pressure and scale compared to AGNs and the active role of pair creation turn out not to alter this process. The simplest theoretical derivation of MRI assumes constant pressure \cite{balbus1998instability}[Eqs. (106)-(110)]; and, while positronium particles disappear and regenerate in flight, overall momentum exchange between radiation and particles preserves gravitational flow. 

Viscous flow and MRI are additive. We interpret $R_{0}$ as the radius where MRI first dominates, giving $V_{MRI} = 0.01(MG/R_{0})^{1/2} \approx (\nu/R_{0})$ \cite{balbus1998instability}[Sect. IV.D], with collisional viscosity $\nu =(C_{s}^{2}\gamma_{c}/(\gamma_{c}^{2} + \omega_{c}^{2})) \leq  (C_{s}^{2}/\gamma_{c})$ for relativistic collision frequency $\gamma_{c}$ and cyclotron frequency $\omega_{c}$. This requires $M > 10^{14}$ g, consistent with the mass range fitting background gamma radiation: where $C_{s}^{2} = (T/m\gamma_{l})$; and $\gamma_{c} = (\rho/m\gamma_{l}^{3}4 \times 10^{8} T_{keV}^{3/2})$. \\


\textbf{4. Gravitational Collapse}

We note that steady flow $\nabla \cdot(n\textbf{v} )= 0$ assumed in Equations (\ref{eq:1a}, \ref{eq:1b}) is only
approximate. The condition to drop $\partial n/\partial t$ in ($\partial n/\partial t + \nabla \cdot (n\textbf{v})$) is ($-v_{r}t > r$). That this
already fails near $R_{0}$ for pure MRI follows from ($-v_{MRI}t) = 0.01 v_{K}t \approx 10^{-5}R_{0} < <R_{0} $ by Equation (\ref{eq:1c}). The resolution is gravitational collapse at $r < R_{0}$ \cite{toomre1964gravitational}.

Collapse can be separated into two stages -- steady flow at constant average
n and T, followed by spherical collapse onto shells of mass $\delta M$
self-confined by their
own gravity ($k_{r} \delta M G$) but still attracted to the black hole. Both stages can be treated
as a gravitational quasi-spherical wave (spherical radius r) coupled to cylindrical
MRI, as follows:

\begin{subequations}
\begin{align}
    \partial n/\partial t + \nabla \textcolor{red}{\cdot}n\textbf{v} = -n\gamma_{A} \label{eq: 2a}\\
    m^{*}\partial/\partial t(n\textbf{v}) = -nm^{*}\nabla(\Phi_{G} + v^{2}) - \nabla p ;  m^{*} = m\gamma_{l} \label{eq: 2b}\\
    m^{*}\partial/\partial t(\nabla  \cdot n\textbf{v}) = -m^{*}\partial/\partial t(\partial n/\partial t + n\gamma_{A}) = \nabla \textcolor{red}{\cdot} [-nm^{*}\nabla (\Phi_{G} + v^{2}) - \nabla p] \label{eq: 2c}\\
    -\partial ^{2}n/\partial ^{2}t \approx n \partial ^{2}/\partial ^{2}r[(MG/r) + v^{2}] - \partial ^{2}(p/m^{*})/\partial ^{2}r - \partial n\gamma_{A}/\partial t \label{eq: 2d}\\
    (\omega + \frac{1}{2}i\gamma_{A})^{2} \approx (v_A^2k_{z}^{2} - 3\Omega^{2}) + (k_{r}^2c_{s}^{2} - \frac{1}{2}k_{r}^{3}c^{2}R_{s}) + (i\gamma_{A}/2)^{2} \label{eq: 2e}\\
    |v_{r}| \approx \Sigma_{k}a_{k}(\omega/k)(k\xi)^{2} \approx |v_{r-MRI}| + (k_{r}^{2}\xi_{r}^{2})c_{s} \label{eq: 2f}\\
    \textbf{E} + c^{-1}\textbf{v}\times\textbf{B} = -<c^{-1}\textbf{v}_{1}\times\textbf{B}_{1}> + \eta\textbf{j} \label{eq: 2g}\\
    \partial \textbf{B}/\partial t = -c\nabla \times \textbf{E} = \nabla \times [\textbf{v} \times \textbf{B} \textbf{+} <\textbf{v}_{1} \times \textbf{B}_{1}> - c\eta\textbf{j}] \label{eq: 2h}
\end{align}
\end{subequations}
Here \textbf{x}, t are dynamical variables; $\Phi_{G}$ is the spherical gravitational potential; and $\gamma_{A}$ represents net annihilation ($\gamma_{A} = 0$ for balanced pair creation and annihilation in the
primordial medium at t$<$14s \cite{weinberg1979first}). As already noted, exchange of momentum
between radiation and particles conserves average gravitational flow.

Equation (\ref{eq: 2c}) applies the divergence operator to Equation (\ref{eq: 2b}), then uses
Equation (\ref{eq: 2a}) to obtain Equation (\ref{eq: 2d}). Equation (\ref{eq: 2e}) approximates Equation (\ref{eq: 2d})by
a representative frequency $\omega$, cylindrical wave number $k_{z}$ and spherical $k_{r}$ in a wave
packet $\propto \Sigma_{k} \exp\  i(k_{r}r + k_{z}z - \omega t)$ yielding the flow in Equation (\ref{eq: 2f}), derived from
Ohm’s Law in Equation (\ref{eq: 2g}) with velocity and magnetic perturbations $\textbf{v}_1$ and $\textbf{B}_{1}$.

Equation (\ref{eq: 2e}) includes both pure MRI driven by rotation $\Omega$ \cite{balbus1998instability} and MRI
Alfv\'en waves coupled to Jeans gravity-driven sound waves \cite{toomre1964gravitational}. This spectrum of
modes is included in wave packet Equation (\ref{eq: 2f}). We omit magnetic interchange,
observed in \cite{mckinney2012general} but weakly additive to pressure here. Waves give $v_{r} = (\omega/k_{r})$, derived
from Ohm’s Law. Ohm’s Law also gives Equation (\ref{eq: 2h}) showing why MRI can initiate
accretion at $R_{0}>>R_{s} = \sqrt{(MG/c^{2})}$. This equation describes the outward propagation
of the MRI dynamo magnetic field, whereby strong MRI near the jet radius
propagates to drive current at $R = R_{0}$ that serves as the O-point around which
polioidal flux circulates -- a phenomenon analogous to kink-driven current drive in
laboratory spheromaks \cite{colgate2014quasi}[App. B]. That Equation (\ref{eq: 2h}) yields the scaling of $R_{0} \propto M^{1/3}t^{2/3}$ is shown in \cite{colgate2015quasi}[Sect. 6.1].

We assume that annihilation controls, serving to adjust perturbations $\xi_{r}$ to maintain $v_{r}r^{2}$ constant in Equation (\ref{eq: 2f}). Constant $v_{r}r^{2}$ preserves $\nabla \cdot(n\textbf{v}) = 0$ giving
constant n and T at their primordial values; hence $\gamma_{A} = 0$ \cite{weinberg1979first}. This condition finally fails
when accretion flow attains its maximum velocity $v_{r} \rightarrow c_{s}$, occurring at a spherical
radius $r = r_{1}$ given by $R_{0}^{2} v_{MRI} \approx r_{1}^{2} c_{s}$. At fixed flow $v_{r} = c_{s}$ inside $r_{1}$, spherical
convergence forces n to grow in competition with radiation and annihilation. It is
this that gives a wave structure $k_{r}r>>1$ representing collapse into thin shells of
mass $\delta M$, too thin to annihilate, with $\gamma_{A} \approx A^{*}n\sigma_{t}c(3/16)(\gamma_{l}^{-2}ln7\gamma_{l})$ with $A^{*} \leq 1$ \cite{krolik1999active, svensson1982pair}[Thomson $\sigma_{t}]$. For large $\gamma_{A}$ we obtain:

\begin{subequations}
\begin{align}
    [\omega + i(\gamma_{A}/2)]^{2} \approx -k_{r}^{3}c^{2}R_{s} - (\gamma_{A}/2)^{2} \label{eq: 3a}\\
    \omega = -i(\gamma_{A}/2) + i(\gamma_{A}/2)[1 + (k_{r}^{3}c^{2}R_{s}/(\gamma_{A}/2)^{2})]^{1/2} \approx i(k_{r}^{3}c^{2}R_{s}/\gamma_{A}) \label{eq: 3b}\\
    (\omega t/k_{r}) > r \rightarrow k_{r}R_{s} > [\gamma_{A}rR_{s}/c^{2}t]^{1/2} \label{eq: 3c} 
\end{align}
\end{subequations}

As in Equation (\ref{eq: 2e}), $\omega$ and $k_{r}$ are representative values in a Fourier analysis of the
shell structure described above. Equation (\ref{eq: 3a}) has two solutions, $\omega = -i\gamma_{A}t$ if
annihilation dominates, or Equation (\ref{eq: 3b}), giving Equation (\ref{eq: 3c}) as the condition on $k_{r}$ for gravity to compete with annihilation $k_{r}R_{s} > 10^{-4}$.

Details do not matter as long as an accretion solution does persist to the
black hole. That some $k_{r}$ can insure this follows from Equation (\ref{eq: 3c}). The required $k_{r}$ is largest where $\gamma_{A}r$ is maximum. The limits on $k_{r}$ are mode coupling (equipartition)
and the magnetic field. The field is unimportant at short wavelengths where $(k_{r})^{-1} <<$ Larmor radius. Only coupling to small $k_{r}$ modes (flattening gradients) would defeat
Equation (\ref{eq: 3c}), not likely here in the time available, due to competition of faster
growth at larger $k_{r}$. Examples of fast turbulence overcoming dynamo equipartition
appear in simulations \cite{archontis2007nonlinear}. Competing with growth is quasi-linear saturation that
acts to reduce the growth rate \cite{krall1973principles}, the net result of growth and mode coupling
giving $k_{r}$ hovering around the instability threshold at $k_{r} \delta M G \rightarrow c_{s}^{2}$.

Absent mode coupling to larger wavelengths, wave pressure does not impede
flow. Pressure is balanced locally inside wave peaks. In fact, instability driving $V_{r} = \omega/k_{r}$ requires that gravity exceed pressure, any excess gravitational force being
balanced by $\partial ^{2}n/\partial ^{2}t$, 
as we assumed in deriving instability Equation (\ref{eq: 3a}).
How $v=(\omega/k)$ 
is obtained from Ohm’s Law, Equation (\ref{eq: 2g}), follows from $\textbf{v} = -[<c^{-1} (\textbf{v}_1 \times \textbf{B}_{1})c\times \textbf{B}/B^{2}] \approx (kv_{1}^{2}/\omega) \approx \omega/k$. 
That this v transports mass in the time available follows from $vt > r$ in Equation (\ref{eq: 3c}). That $v \leq c$ follows from $r_{1}/t<c$, and energy is conserved. So no physics is violated. That v is independent of T means that accretion proceeds with or without severe cooling, the maximum annihilation rate being $\approx n\sigma_{t}c$ also independent of T \cite{krolik1999active}[Eq. (8.47)]. \\

\textbf{5. Summary}

We have found a possible path for MRI-driven accretion creating black holes
that could explain dark matter abundance. The regime of interest is defined by
matching background gamma radiation to Hawking radiation, yielding $M = 10^{15-18}$ gm as the mass range for dark matter black holes. A rotating plasma suppressing Bondi
accretion and stimulating MRI can produce masses in this range.

As already noted, accretion by MRI is an extension of gravitational collapse
creating a seed mass. The main role of MRI is to propagate the accretion radius far
from the black hole. That MRI does this around AGNs is inferred from evidence for
strong magnetic fields near the black hole \cite{zamaninasab2014dynamically} and other evidence for magnetic
fields far from the jet axis \cite{colgate2015quasi}. Theoretically, both for AGNs and the primordial
plasma, propagation of the magnetic field to its O-point is accounted for by
turbulence in Ohm’s law independent of pressure.

It is slow MRI flow at the O-point that produces pressure gradients relieved
by collapse. Collapse occurs in two stages. The first stage is a perturbation on quasi-
steady incompressible flow that does not much disturb the primordial medium,
from $r = R_{0}$ down to the radius $r_{1} \approx 10^{-5}R_{0}$ mentioned above. The other critical
dimension is the MRI jet radius $a \approx (C^{2}\eta t/4\pi)^{1/2}) \approx 0.1 t^{7/8} < 10^{-5}R_{0}$ determined by
collisional resistivity $\eta$.

The second stage of collapse joins steady flow at $r = a$ to Jeans gravitational
instability, self-adjusted to outrun annihilation if the temperature cools along
accretion paths. In both stages accretion flow is sustained by unstable accretion
waves. As noted above, the fact that strong instability is required in the final stage is
a natural extension of Jeans gravitational collapse creating the seed mass \cite{toomre1964gravitational}. \\

\textbf{{6. Concluding points}}

Gravity couples MRI Alfven waves to Jeans sound waves
$\propto$ exp i($k_{r}r$). The
full dispersion relation then has the form $\omega^{4} - C\omega^{2} + D = 0$ \cite{balbus1998instability}[Eq. (111)], with
additional terms $\propto i\omega^{3}$, $i\omega$ to account for resistivity and viscosity \cite{balbus1998instability}[Eqs. (135-138)]. Adding gravity adds $H^{*}$ to both C and D, where $H^{*} = (k_{r}^{2}c_{s}^{2} - \frac{1}{2}k_{r}^{3}c^{2}R_{s})$ appearing
in the approximate Equation (\ref{eq: 2e}).

Primordial MRI accretion is due to dynamo space charge, $\sigma = \nabla . (\textbf{E}/4\pi)$, giving as the jet momentum equation $d\textbf{P}/dt = c^{-1}\textbf{j}^{*} \times \textbf{B} = c^{-1}\textbf{j} \times \textbf{B} + \sigma \textbf{E}$ \cite{fowler2019quasi}[Eq. (38b)]. Here $\textbf{j}^{*}$ is a short-circuit current circulating near the black hole \cite{fowler2019quasi}[Fig. 4]. It is this short circuit that drives spherical accretion in Equation (\ref{eq:1a}). Numbers in
Equation (\ref{eq:1a}) are: density $\rho = nm\gamma_{l} \propto T^{4}$ \cite{weinberg1979first}[p.158] with ambient $T_{amb} = 10^{10}$ $^{\circ} K/\sqrt{t}$ fitting the range $T = 10^{9-11}$ $^{\circ} K$ \cite{weinberg1979first}[pp. 4,5] and Lorentz $\gamma_{l} \approx [1 + (T/mc^{2})]$. Then $\rho = (X 10^{5}/t^{2})$ g/cc, with X = 4 at t = 0.01 s \cite{weinberg1979first}[p.4] down to X = 1.22 after annihilation \cite{weinberg1979first}[p.159], giving $f^{*}$ independent of M for all M’s within our mass range above.

Magnetic jets are reviewed in \cite{blandford2019relativistic}. For the highly collisional primordial
plasma, a crucial point is that MRI does not depend on cyclotron orbit closure. What
does matter is resistive and viscous diffusion in competition with instability growth
rates \cite{balbus1998instability}[Sect. IV D], which we take into account in criteria such as requiring viscous $k_{r}^{2}\nu < \Omega$, and the dynamo failure when $\nu/R > v_{MRI}$.

Another issue is turbulence fluctuation levels. As for AGN’s, we assume that
fluctuation levels in Equation (\ref{eq: 2f}) always adjust to sustain accretion, since
otherwise there could be no MRI power source. We found that fluctuation $\xi/R \approx 0.1$ explained dark matter abundance. A demonstration of these turbulence levels
awaits simulations of the primordial plasma on general relativistic codes like that
in \cite{mckinney2012general}; or special relativistic codes treating the black hole as a conducting sphere \cite{tchekhovskoy2016three}. The run times for GRMHD codes, sometimes too short to produce a fully-developed jet tower \cite{fowler2019quasi}[App. A], may be better suited to the primordial regime.

The primordial magnetic field order 1 gauss could have been created by
Biermann battery action due to pressure fluctuations in the ordinary matter content
of primordial plasma (though positronium contributions tend to cancel) \cite{kulsrud1997protogalactic}.The only source of rotation is the Peebles torque coupling neighboring black hole domains, giving $\int dx$  $\partial/\partial t$ ($\varrho R(MG/R)^{1/2}) = \alpha(M^{2}G/R)$ for elliptic distortion $\alpha$ \cite{peebles1969origin}. Other terms (MRI jet, viscosity) only recycle angular momentum \cite{balbus1998instability}, Eq. (29). The Peebles source is only essential near peak MRI activity around the jet radius R = a, sufficient for $M > $ ($\int dx$ $\partial/\partial t$  $[(\varrho R(G/R)^{1/2}/\alpha G])^{2/3}  \approx 10^{6}(a^{7/3}/\alpha^{2/3})$. As already noted, MRI current drive propagates from R = a to R = $R_0$.

Fitting our mass range $M = 10^{15-18}$ g to observed background gamma rays
comes from:

\begin{subequations}
\begin{align}
     Eg(E) = \Sigma_{i}N_{i}M_{i}(R_{si}/<R>)^{2}(\sigma_{SB}T^{3}_{Hi}/kM_{i})P_{Planck}\label{eq: 4a}\\
     = \int_{M1}^{A2/E} dM (f(M)/M_{DM})(A_{1}/M)[(EM/A_{2} - (EM/A_{2})^{2}]\label{eq: 4b}\\
     \int_{M1}^{M2} dM M f(M) \approx (M_{2}/0.5 \times 10^{18})^{3}M_{DM} \label{eq: 4c}
\end{align}
\end{subequations}
where $M_{1} = 10^{15}$ g and $g(E) = 0.1(10keV/E)^{2}$ fits data in \cite{trombka1983gamma}. Equation (\ref{eq: 4a}) describes Hawking radiation (temperature $T_H\propto 1/M $, Planck distribution) from $N_{i}$ black holes each of mass $M_{i}$. The result is about the same for universal and local signals, giving $\Sigma_{i}N_{i}M_{i}(R_{si}/<R>)^{2} = M_{DM}/<R>^2 = 0.03 - 0.05$ for average radius $<R> = 3 \times 10^{23}$ and $M_{DM} = 3 \times 10^{45}$g for the Milky Way and $<R> = 4.4 \times 10^{28}$ and $M_{DM} = 10^{56}$g for the visible Universe. Approximating the sum by a mass distribution f(M) gives Equation (\ref{eq: 4b}) as
an integral equation for f(M) with approximate solution $f(M) = (20/4\pi A_{1} A_{2} )M_{DM} M$ with $A_{1} = 5.4 \times 10^{32}$ and $A_{2} = 7.5 \times 10^{19}$. The magnitude of f(M) is determined, giving
Equation (\ref{eq: 4c}) showing that the total mass of black holes approximately equals the
total mass of dark matter, for $M_{2} \approx 10^{18}$. \\

\textbf{Acknowledgements} \\
\indent TKF thanks Christopher McKee and Hui Li for many enlightening discussion. TKF and RA also thank Nathan Ngata for timely aid in  editing the manuscript.

\end{document}